\documentclass[10pt,conference]{IEEEtran}

\usepackage{cite}
\usepackage{amsmath,amssymb,amsfonts}
\usepackage{algorithmic}
\usepackage{enumitem}
\usepackage{graphicx}
\usepackage{textcomp}
\usepackage[dvipsnames]{xcolor}
\usepackage{xspace}
\usepackage{booktabs}
\usepackage{multirow}
\usepackage{todonotes}
\usepackage{hyperref}
\usepackage{balance}

\usepackage[most]{tcolorbox}
\usepackage{enumitem}

\newif\ifshowcomments
\showcommentstrue % turn ON - comment this line to turn them off

\begin{document}

\title{Early Reflections on How Large Language Models are Reshaping Software Engineering Research}

\author{\IEEEauthorblockN{David Williams}
\IEEEauthorblockA{\textit{University College London}\\
London, United Kingdom \\
david.williams.22@ucl.ac.uk}
\and
\IEEEauthorblockN{Michail Kyrkos}
\IEEEauthorblockA{\textit{Athens University of Economics and Business}\\
Athens, Greece \\
mic.kyrkos@aueb.gr}
\and
\IEEEauthorblockN{Max Hort}
\IEEEauthorblockA{\textit{Norsk Regnesentral}\\
Oslo, Norway \\
mhort@nr.no}
\and
\IEEEauthorblockN{Maria Kechagia}
\IEEEauthorblockA{\textit{National and Kapodistrian University of Athens}\\
Athens, Greece \\
makechag@ba.uoa.gr}
\and
\IEEEauthorblockN{Justyna Petke}
\IEEEauthorblockA{\textit{University College London}\\
London, United Kingdom \\
j.petke@ucl.ac.uk}
\and
\IEEEauthorblockN{Federica Sarro}
\IEEEauthorblockA{\textit{University College London}\\
London, United Kingdom \\
f.sarro@ucl.ac.uk}
}

\maketitle

\begin{abstract}

Large language models (LLMs) are radically reshaping software engineering (SE) research. Yet, their steep financial and computational costs raise questions about who participates in that change, and who does not. By analysing all 1,918 papers published in the ICSE 2019--2026 Research and SEIP tracks, we uncover a shifting and highly concentrated research landscape. China and the US dominate LLM-based SE research, appearing in 74\% of LLM-based publications between 2023--2026, yet they diverge sharply in their use of commercial versus open-weight models. Beneath this national dominance lies extreme institutional concentration: just 5 institutions in China and 7 in the US appear in more than half of their respective nations' LLM-based output. In the US, only half of the unique institutions active in 2023--2026 participated in LLM-based research, whereas Singapore displays broader usage, with 11 of 15 active institutions participating in LLM-based work. Furthermore, across 2023--2026 Research-track papers, industry involvement is more prevalent in LLM-based work (23.4\% vs 16.9\%), and academia-only papers make up a smaller share (63.4\% vs 70.9\%). Crucially, we expose a gap in model usage across industry and academia. SEIP track papers lean strongly towards commercial LLMs, with 77.4\% featuring them and 40.5\% using them exclusively. In contrast, only 21.2\% of Research-track work relies strictly on commercial models, making greater use of open-weight models that typically support more sustainable and reproducible experiments. However, this divide warrants reflection on the extent to which SE research aligns with industry practice.

\end{abstract}


\section{Introduction}

It is no secret that Large Language Models (LLMs) have come to dominate the field of Software Engineering (SE) research in recent years. However, the substantial financial and environmental costs of LLM-based SE research remain an ``elephant in the room'' within the community.

These costs risk contributing to geographical and institutional disparities in access to the computational resources required to participate in research~\cite{khandelwal2024100}. As early as 2021, industry-developed AI models were, on average, 29 times larger than those developed in academia~\cite{ahmed2023growing}. More recently, a 2025 questionnaire of 57 authors of LLM-based ICSE papers found that more than half were uncertain about their ability to sustain the costs of experiments involving commercial LLMs~\cite{williams2026nier}. Together, these observations raise concerns about the continued accessibility of SE research, echoing findings from over a decade ago that just 7\% of countries accounted for 60\% of SE publications between 1969 and 2009~\cite{garousi2013geographic}.

Existing work on LLM-based research focuses on publication trends~\cite{fan2024bibliometric, movva2024topics, xia2025analyzing}, LLM applications across different SE tasks~\cite{fan2023survey,hou2024slr}, the geographical distribution of research activity, industry involvement, and the adoption of commercial LLMs~\cite{WBY27}, as well as issues related to sustainability~\cite{williams2026nier,shi2024green} and reproducibility~\cite{angermeir2026reproducibility}. Beyond LLMs, bibliometric studies of SE have examined leading institutions, researchers, research topics, geographical distributions, citation patterns, and funding sources~\cite{karanatsiou2019bibliometric,wong2021bibliometric,garousi2016citations,kapitsaki2026does}. Our study (1) compares LLM-based and non-LLM-based SE research, (2) investigates collaboration patterns among countries and institutions, and (3) explores differences in how countries and institutions use commercial and open-weight LLMs. Collectively, our findings highlight the need for more accessible LLMs and equitable research opportunities across countries and institutions.

We conducted an empirical study of all 1,918 papers published in the ICSE Research and SEIP tracks between 2019 and 2026, combining keyword filtering, LLM-assisted classification, and manual validation to identify 488 LLM-based papers. We extracted institutional affiliations and countries for all examined papers and classified LLM-based studies as using open-weight models, commercial models, or both. In analysing the global distribution of top-tier SE research, we set out to answer the following:

\textbf{RQ1. What is the LLM adoption in SE research across countries and institutions?}
We identify the leading countries and institutions in LLM-based and non-LLM SE research. We also investigate which countries and institutions contribute proportionately more LLM-based than non-LLM-based research.

\textbf{RQ2. How do countries and institutions use open-weight and commercial models?} Our answer explores whether countries have differing patterns in commercial and open model use, as well as the models used by newcomers to the field.

\textbf{RQ3. What are the institutional and international collaboration patterns for LLM- and non-LLM-based research?}
We compare collaborations among countries and institutions in LLM-based research with pre-LLM research (2019--2022) and non-LLM research in the post-LLM era (2023--2026).

\section{Methodology} \label{sec:methodology}

\begin{figure}
    \centering
    \includegraphics[width=0.9\linewidth]{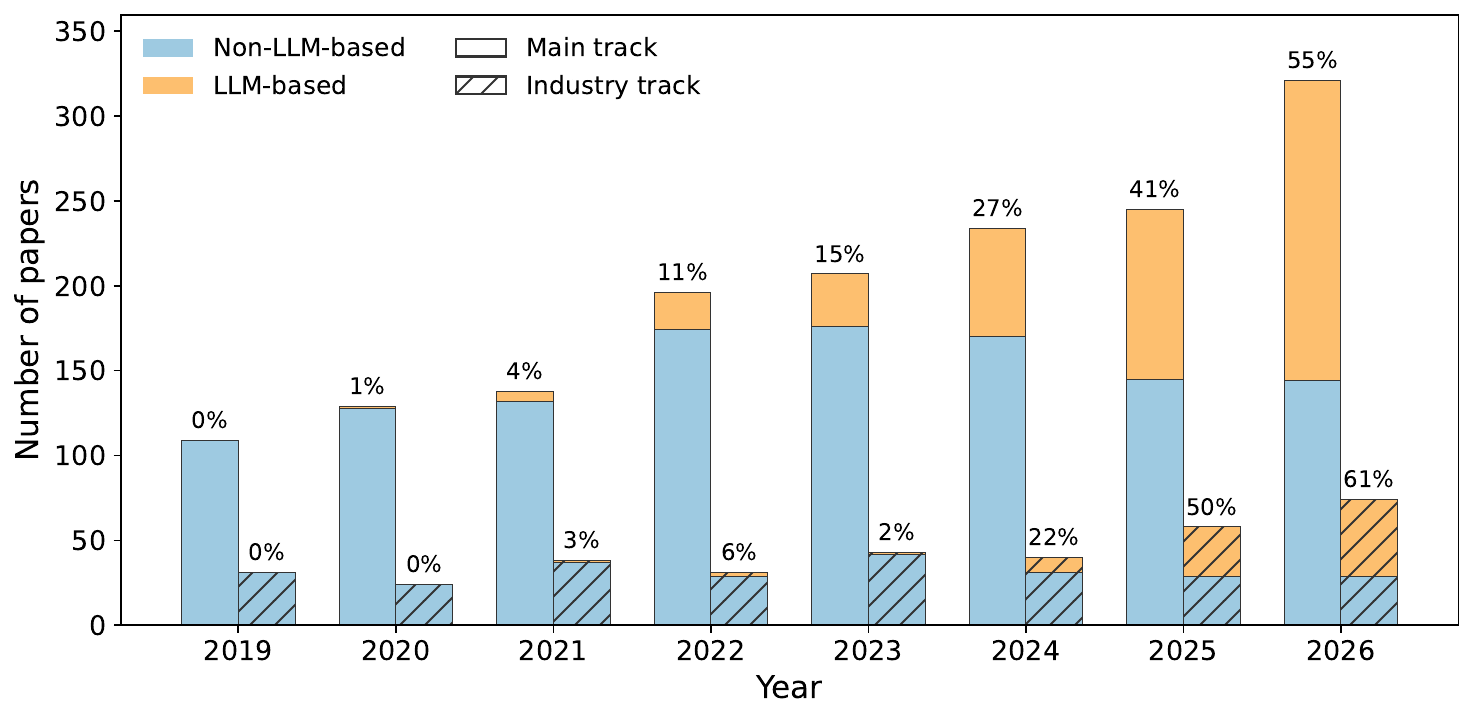}
    \caption{Distribution of LLM-based and non-LLM-based papers published at ICSE from 2019--2026. Labels on each bar indicate the percentage of research that was LLM-based for each year (or transformer-based for earlier years).}
    \label{fig:paper_distribution}
\end{figure}

We gathered, extracted data from, and analysed papers published in the ICSE Research and Software Engineering in Practice (SEIP)\footnote{We selected SEIP to analyse industry involvement in SE research.} tracks from 2019 to 2026, representing a four-year period before the mass proliferation of LLMs (2019--2022, coinciding with the late 2022 release of ChatGPT), and the four latest years of research (2023--2026). In line with prior work~\cite{williams2026nier}, we focus on papers published at ICSE, the premier venue in SE that garners the most international attention. In total, our dataset includes 1,918 papers: 1,579 from the Research track and 339 from SEIP. Figure~\ref{fig:paper_distribution} shows papers by year, split into LLM-based and non-LLM-based work.

\textbf{Keyword and LLM-Based Paper Filtering.} \label{sec:methodology_filtering}
To analyse differences between LLM- and non-LLM-based research, we first identified all papers that evaluated, applied, or investigated LLMs in an experimental setting (i.e., including user studies but excluding purely literary works such as surveys and mapping studies). Starting with the 1,918 papers published in the Research and SEIP tracks of ICSE 2019--2026, we used a hybrid filtering process combining keyword-based search with automated LLM classification, following prior work~\cite{BSO24, KPK24, LTM25, SHJ25, ASL26}.

For keyword filtering, we searched titles and abstracts using keywords from Williams et al.~\cite{williams2026nier}, namely \texttt{LLM}, \texttt{language model}, \texttt{conversational AI}, \texttt{chatbot}, \texttt{genAI}, \texttt{generative}, \texttt{AI-assisted}, \texttt{foundation model}, \texttt{BERT}, \texttt{encoder}, \texttt{decoder}, \texttt{autoencoder}, \texttt{transformer}, and \texttt{agent} (originally adapted from the LLM definition provided by Fan et~al.~\cite{fan2023survey}), yielding 499 papers labelled as potentially LLM-based.

To improve classification accuracy from keyword matching, we classified all 1,918 titles and abstracts using two independent LLMs: \texttt{GPT-6 Luna} (9/10/26) and \texttt{Claude Sonnet 4.6} (9/10/26).\footnote{Both LLMs were configured with temperature at zero, a max. output length of 1,024 tokens, and a 60-second timeout. We disabled reasoning effort for \texttt{gpt-6-luna} and thinking for \texttt{claude-sonnet-4.6}.} With the paper's title and abstract as context, each model was prompted to classify whether the research is LLM-based (with options \textit{yes}, \textit{no}, or \textit{unclear}), and provide verbatim quotes from the title or abstract justifying its verdict. Across the LLMs, we found only 80 disagreements (95.8\% agreement; Cohen's $\kappa$=0.889), with more than half (53.8\%) occurring when one LLM labelled \textit{unclear} while the other gave a definitive label.

To establish the final list of LLM- and non-LLM-based papers, we manually resolved cases where any of the three label sources (keywords, \texttt{GPT-6 Luna}, \texttt{Claude Sonnet 4.6}) disagreed. In total, we resolved 171 (8.9\%) paper labels. Of the 488 LLM-based papers, 456 (372 Research, 84 SEIP) appeared in 2023–2026, with the remaining 32 being transformer-based works from 2019–2022. Our replication package~\cite{repo} provides a complete audit of our labelling process, including prompts and model configurations.

\textbf{Extracting Countries and Institutions.}
First, we extracted the authors' institutional affiliations for each of the 1,918 papers. To do so, we used each paper's DOI to retrieve affiliation information from the OpenAlex~\cite{openalex_api} and CrossRef~\cite {crossref_api} APIs. OpenAlex provided affiliation information for 1,915 papers, while Crossref provided information for 1,656 papers. 

From each affiliation, one reviewer manually extracted three pieces of information: the country, the standardised institution name (ignoring departments or research groups), and the institution type. Institution type is either \textit{academic} (for educational institutions, e.g., University of Washington), \textit{industry} (for companies, e.g., Microsoft), or \textit{other} (for independent research institutions, e.g., Alan Turing Institute). The OpenAlex and Crossref APIs did not provide a country for 98 papers. To label these, we manually checked several sources, including the paper's first page, Google search, and author LinkedIn profiles. When a paper associated the same institution with different countries, we retained a separate entry for each country.

Combining this affiliation dataset with our papers yields a list of all papers published at ICSE in the Research and SEIP tracks from 2019--2026, annotated with each participating institution's standardised name, type, and country.

\textbf{Extracting Model Types.}
To investigate the types of LLMs used in SE research, we analysed the full text of all 488 LLM-based papers. First, we downloaded the {\sc PDF} of each paper and extracted its textual content (excluding references) using the {\tt PyMuPDF} library~\cite{pymupdf}. Second, we used two LLMs, \texttt{gpt-6-luna} (21/10/26) and \texttt{gemini-3.8-flash} (21/10/26), to independently classify each paper according to the types of models used.\footnote{Both LLMs were configured with temperature at zero, a max. output length of 16,384 tokens, and a 120-second timeout. We disabled reasoning effort for \texttt{gpt-6-luna} and set thinking level to low for \texttt{gemini-3.8-flash}.} Given the text, each LLM was instructed to assign one of four labels: \textit{open} (e.g. BERT, StarCoder, CodeLlama), \textit{commercial} (e.g. OpenAI's GPT, Anthropic's Claude), \textit{both}, or \textit{unclear}.

For 84 of the 488 papers, the models either disagreed (77 papers) or both returned \textit{unclear} (7 papers). These 84 papers were manually examined by one reviewer by looking at the full text and assigning the final labels. To assess the labelling accuracy for the 404 papers where the models agreed, we manually inspected 81 papers (20\%). This inspection revealed only three misclassifications (3.7\% error rate), which we corrected: We reclassified two papers initially labelled \textit{commercial} as \textit{both}, and one initially labelled \textit{open} as \textit{both}.

\begin{table}[t!]
    \caption{RQ1: Top ten countries by number of papers published at ICSE. A paper is counted once for each represented country.}
    \label{tab:top-countries}
    \centering
    \resizebox{\linewidth}{!}{
    \begin{tabular}{rlll}
        \toprule
         & & \multicolumn{2}{c}{2023--2026}\\
         \cmidrule{3-4}
         Rank & 2019–2022 (n=696) & Non-LLM-based (n=766) & LLM-based (n=456) \\
        \midrule
        1 & US (48\%) & China (39\%) & China (45\%) \\
        2 & China (32\%) & US (36\%) & US (32\%) \\
        3 & Australia (11\%) & Canada (12\%) & Singapore (14\%) \\
        4 & Canada (10\%) & Germany (10\%) & Canada (12\%) \\
        5 & Germany (8\%) & Singapore (7\%) & Germany (11\%) \\
        6 & Singapore (7\%) & Hong Kong (7\%) & Australia (9\%) \\
        7 & Switzerland (6\%) & Australia (5\%) & Hong Kong (6\%) \\
        8 & UK (6\%) & UK (5\%) & UK (5\%) \\
        9 & Hong Kong (4\%) & Switzerland (4\%) & Switzerland (3\%) \\
        10 & South Korea (4\%) & Netherlands (3\%) & South Korea (3\%) \\
        \bottomrule
    \end{tabular}
    }
\end{table}
\section{Results}
In the following sections, we present our findings on the global distribution of SE research. We use the terms ``pre-LLM'' and ``post-LLM'' throughout to refer to the 2019--2022 and 2023--2026 periods, respectively.

\subsection{RQ1. LLM Adoption Across Countries and Institutions}

To explore country participation, we examined the number of unique countries represented in each time period. Across the Research and SEIP tracks, the number of active countries increased from 44 to 53 between the pre- and post-LLM periods. However, while the number of countries involved in non-LLM-based work increased to 47 in the post-LLM period, only 38 countries participated in LLM-based research. Among the 15 countries active in the post-LLM period but not in LLM-based research, six were from Europe, four from South America, three from Asia, and two from Africa.

Table~\ref{tab:top-countries} shows the top contributing countries across the pre- and post-LLM periods and, within the post-LLM period, across LLM- and non-LLM-based research. In 2016, Garousi et al.~\cite{garousi2016citations} showed that 3 countries (US, China, and the UK) produced almost half of all SE papers published between 1972 and 2013. In the post-LLM period, Chinese institutions alone are represented in 45\% of all LLM-based SE research at ICSE, with the US following at 32\% and the emergence of a third player, Singapore, at 14\%. Together, Chinese and US institutions appear in 73.9\% (n=337) of LLM-based papers in 2023--2026, aligning with recent findings that these nations are among the largest funders of SE research~\cite{kapitsaki2026does}.

Among the top 10 countries, we observe notable differences in their representation across LLM- and non-LLM-based works. For instance, while Chinese institutions appear in 39\% of non-LLM-based research, they appear in 45\% of LLM-based work. Australian institutions also feature in proportionally more LLM-based papers (5\% to 9\%). This gap is even more extreme for Singapore, which contributes to only 7\% of non-LLM-based research but doubles its presence in LLM-based papers (14\%). In contrast, the United States drops from contributing 36\% of non-LLM-based research to 32\% of LLM-based works. Some countries, such as Canada and the United Kingdom, retain balanced representation across LLM- and non-LLM-based works (12\% and 5\% in both periods, respectively), while the Netherlands drops out of the top 10, replaced by South Korea.

Looking at institutions for the major players in LLM-based SE research, we see that in China, five institutions appear on more than 50\% of the 207 LLM-based papers authored by 82 institutions in the post-LLM era. Similarly, in the US, while 165 institutions participated in SE research in the post-LLM era, only 85 are involved in LLM-based research, and as few as 7 institutions feature in more than half of that LLM-based work. In contrast, for Singapore, 11/15 institutions active in the post-LLM era have contributed to LLM-based work.

\subsection{RQ2. Trends in Open and Commercial Models}

For each LLM-based paper in our dataset, we labelled whether it featured open-weight models (e.g. BERT, StarCoder, CodeLlama), commercial ones (e.g. OpenAI's GPT, Anthropic's Claude), or both. Comparing the Research and SEIP tracks in the post-LLM era, we observe striking differences: Across the 84 SEIP papers, 77.4\% feature commercial models, and 40.5\% use exclusively commercial ones. In contrast, the Research track features open models more frequently, with only 21.2\% of post-LLM era papers using exclusively commercial models. This observation is encouraging for the state of research (since SE researchers typically believe that research with open models is more sustainable~\cite{williams2026nier}), but also raises the question of how aligned research and practice are, as it has been shown that task performance does not necessarily translate across models and open-source benchmarks do not represent industrial workloads~\cite{rondon2025agentaprgoogle}. On the other hand, while commercial models typically offer frontier performance, industry reliance on them exposes clients to several challenges, including potential vendor lock-in, data privacy risks, and unpredictable opaque model updates.

To highlight differing model usage patterns across countries, we compare the two top contributors in Table~\ref{tab:top-countries}. Focusing on post-LLM papers authored exclusively by Chinese institutions versus those authored exclusively by US institutions, we find that each country's largest share of LLM-based papers featured both open-weight and commercial models (56.0\% for China, 40.3\% for the US). However, the US had a substantially higher proportion of papers using solely commercial models than China (34.7\% vs only 9.9\%, respectively). For papers featuring exclusively open models, China had proportionally more, at 34.1\% compared with the US at 25.0\%.

Finally, we focus on the latest year of research to understand model usage among newcomers (defined as institutions appearing in 2026 that did not publish in either track at ICSE 2019--2025). Out of 121 newcomers in 2026, 64 contributed to LLM-based research (34 academic, 24 industry, 6 other). Compared to the 2023--2025 baseline, the 59 papers featuring these newcomer institutions incorporated commercial models at a substantially higher rate (81.4\% vs 67.5\%). Further, in 2026, papers featuring newcomers were notably more likely than those featuring only established peers to rely exclusively on commercial models (28.8\% vs 18.4\%).

\subsection{RQ3. Collaboration Patterns}

\begin{figure}
    \centering
    \includegraphics[width=0.9\linewidth]{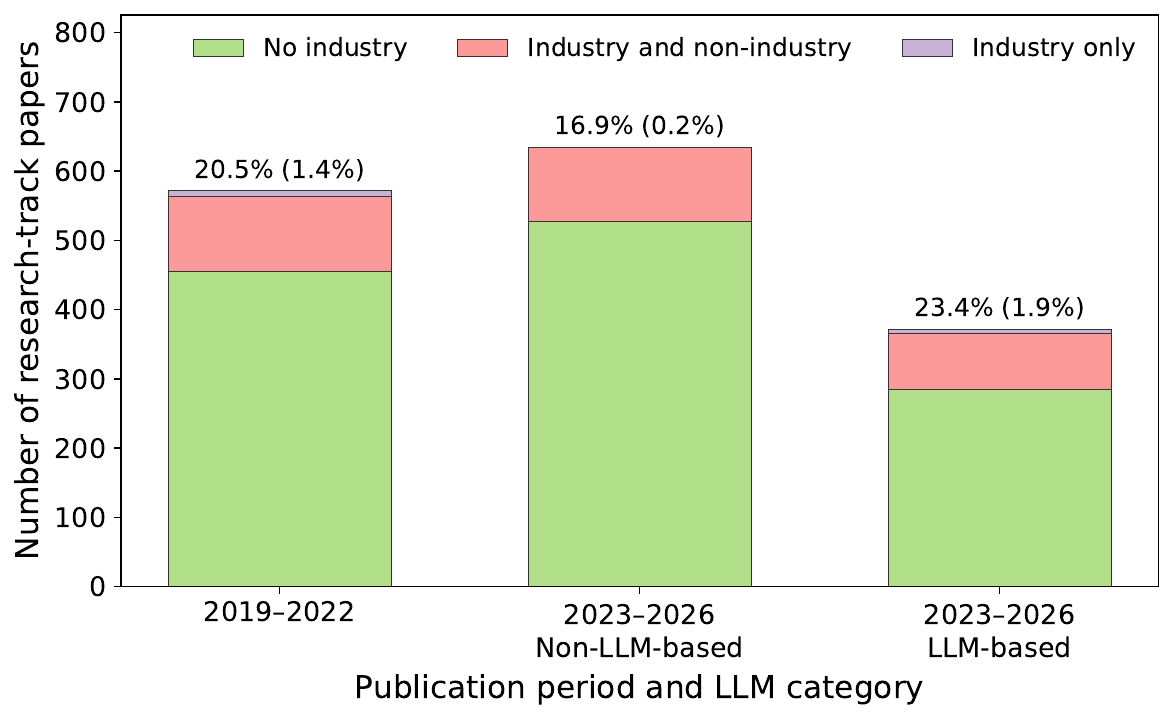}
    \caption{RQ3: Industry involvement across the pre- and post-LLM periods in the ICSE Research track. Percentage labels represent the proportion of papers with industry involvement, with percentages in brackets representing the proportion of papers that feature only industry.}
    \label{fig:industry-involvement}
\end{figure}

Since we explore industry involvement in this RQ, we analyse the Research and SEIP tracks separately.

Figure~\ref{fig:industry-involvement} illustrates the proportion of industry involvement in the ICSE Research track across the pre- and post-LLM periods. Industry focus has slightly shifted toward LLM-based research, from 20.5\% (117/572) involvement in the pre-LLM period to 23.4\% (87/372) for LLM-based works, and down to 16.9\% (107/635) for non-LLM-based works. This pattern continues in the proportion of industry-only papers in the post-LLM period, which rose to 1.9\% for LLM-based papers and dropped to 0.2\% for non-LLM-based papers. In the pre-LLM era, 71.9\% of papers featured only academic institutions; in the post-LLM era, this proportion stayed at 70.9\% for non-LLM-based works but decreased to 63.4\% for LLM-based works. Academic--academic (multiple institutions, all academic) and industry--academia (at least one company and one academic institution) collaborations are common in LLM-based papers (40.6\% and 21.2\%, respectively). Meanwhile, papers involving more than one company remain rare and are no more common than in non-LLM-based work (3.2\% vs 3.1\%; 2.3\% pre-LLM era), following prior observations on LLM-based research~\cite{movva2024topics}.

Across eras, the proportion of papers in the Research track featuring international collaboration (authors from more than one country) has decreased from 48.6\% pre-LLM to 41.7\% post-LLM. In the post-LLM era, 70.9\% of LLM-based academic--academic papers are international, substantially higher than their non-LLM counterparts (61.9\%) and comparable to the 71.1\% rate in the pre-LLM era. In contrast, industry--academia papers exhibit similar rates in LLM- and non-LLM-based work (49.4\% vs 51.5\%), both lower than the 63.0\% rate in the pre-LLM era.

In the SEIP track, we see a major shift toward LLM-based research. Figure~\ref{fig:paper_distribution} shows that the proportion of LLM-based works nearly tripled from 22\% in 2024 to 61\% in 2026. Collaboration patterns here contrast with the Research track. First, SEIP consistently features a large share of industry--academia collaborations, which remained stable at $\sim$58\% across both periods and across LLM- and non-LLM-based works. Second, while industry-only papers are virtually absent in the Research track, they account for 29.8\% of SEIP LLM-based works, 19.1\% of non-LLM-based works and 22.6\% in the pre-LLM period. Finally, only 31.0\% of SEIP LLM-based works feature international collaboration, the lowest of the three groups (36.6\% of post-LLM era non-LLM-based works; 41.9\% in the pre-LLM era), while Research-track LLM-based work remained close to its pre-LLM level (47.8\% vs 48.6\%).

\section{Threats to Validity}

To mitigate the threat of misclassifications (i.e. LLM-based papers labelled as non-LLM-based, or vice versa), we use a hybrid keyword--LLM filtering pipeline and perform manual inspections at several stages. However, we do not manually validate papers on which all three label sources agreed. Consequently, LLM-based papers that all three methods miss may remain undetected. We used LLMs with manual validation to handle the large volume of papers. While this approach aligns with recent advances in literature surveys~\cite{BSO24,KPK24,LTM25,SHJ25,ASL26}, we acknowledge it can pose reproducibility risks~\cite{williams2026nier}. To mitigate this, we followed established guidelines~\cite{baltes2026guidelines} and documented our complete process in our repository~\cite{repo}. Beyond misclassifications, our data covers affiliations and model types, not financial or compute costs, meaning we approximate accessibility rather than measure it. Finally, since research is highly concentrated in a few institutions, country-level findings can largely reflect the behaviour of those institutions.

\section{Conclusions and Future Plans}

We presented a snapshot of how LLMs are reshaping the geographical and institutional boundaries of SE research, analysing 1,918 ICSE Research and SEIP articles. LLM-based research is concentrated in a small number of countries and institutions, with greater industry involvement than non-LLM-based work. Substantial differences in model adoption across countries and tracks warrant reflection on how well academic research aligns with industrial practice. Future work should examine the prominence of LLM-based work across SE sub-disciplines to see whether certain subfields are disproportionately shaped by LLMs. It should also move beyond the binary open-versus-commercial distinction to track model families, parameter sizes, and hardware or token costs, further characterising accessibility. As LLMs become central to SE research, the community should work to ensure that access to them does not determine who gets to shape the field.

\bibliographystyle{IEEEtran}
\bibliography{IEEEabrv,references}

\end{document}